\def\FORMAT{0} 

\ifnum\FORMAT=0
    \def\ARXIV{1} 
    \def\BLIND{0} 
    \def\BBL{0} 
\else
    \ifnum\FORMAT=1
        \def\ARXIV{0}

        \def\BLIND{1}
        \def\BBL{0}
    \else
        \def\ARXIV{0}

        \def\BLIND{0}
        \def\BBL{1}
    \fi
\fi

\documentclass[USenglish,12pt]{article}

\usepackage[utf8]{inputenc}
\usepackage[big,online]{dgruyter}    
\usepackage[T1]{fontenc}             
\usepackage{lmodern}
\usepackage{anyfontsize}             
\usepackage{microtype}
\microtypesetup{patch=none}          
\usepackage{silence}
\usepackage[letterpaper,top=1in,bottom=1in,left=1in,right=1in,headheight=14pt]{geometry}     
\usepackage{setspace}
\usepackage[authoryear,round,sort&compress]{natbib}

\usepackage{amsmath}
\usepackage{amssymb}

\usepackage{booktabs}
\usepackage{multirow}
\usepackage{graphicx}
\usepackage{float}
\usepackage{longtable}
\usepackage{subcaption}

\AtBeginDocument{%
  \renewcommand{\tablename}{Table}%
  \renewcommand{\figurename}{Figure}%
}
\usepackage{threeparttable}
\usepackage{siunitx}
\usepackage{algorithm}
\usepackage{algpseudocode}

\makeatletter
\providecommand{\plist@algorithm}{\plist@figure}
\makeatother

\usepackage[hyphens]{url}
\usepackage{xcolor}
\usepackage{tcolorbox}
\usepackage{xparse}
\usepackage{xspace}
\usepackage[normalem]{ulem}
\robustify\uline
\usepackage{soul}
\useunder{\uline}{\ul}{}
\usepackage{orcidlink}
\usepackage{svg}

\theoremstyle{dgthm}

\theoremstyle{dgdef}

\newcommand{\wbp}{\text{Weibull}_{\psi}}
\newcommand{\wbpc}{\text{Weibull}_{\kappa,\psi}}
\newcommand{\zoup}{\text{Zou}_{\psi}}
\newcommand{\zouc}{\text{Zou}_{\kappa,\psi}}
\newcommand{\maiap}{\text{Maia}_{\psi}}
\newcommand{\maiac}{\text{Maia}_{\kappa,\psi}}
\newcommand{\poik}{\text{Poisson}_{\kappa}}
\newcommand{\poikt}{\text{Poisson}_{\kappa,t}}
\newcommand{\naivek}{\text{Naive}_{\kappa}}

\newcommand{\jc}[1]{}
\newcommand{\lc}[1]{}

\ifnum\ARXIV=1
    \usepackage{hyperref}
    \usepackage{doi}
    
    \usepackage{orcidlink}
    \usepackage{caption}
    \usepackage{lastpage}
    \hypersetup{
        pdftitle={A market-initialised accelerated failure time model for in-play football forecasting},
        pdfauthor={Lawrence Clegg, Zixing Song, John Cartlidge},
        pdfsubject={In-play football forecasting},
        pdfkeywords={betting markets, market calibration, goal arrival models, soccer, survival analysis, Weibull distribution}
    }
    \makeatletter

    \def\ps@plain{%
      \let\@oddhead\@empty
      \let\@evenhead\@empty
      \def\@oddfoot{%
        \footnotesize
        Preprint: Clegg, Song and Cartlidge (2026)%
        \hfill Page \thepage\ of \pageref{LastPage}%
      }
      \let\@evenfoot\@oddfoot
    }

    \def\ps@headings{%
      \def\@oddhead{%
        \footnotesize
        In-play football forecasting%
        \hfill Preprint: Clegg, Song and Cartlidge (2026)%
      }
      \let\@evenhead\@oddhead
      \def\@oddfoot{%
        \footnotesize
        \hfill Page \thepage\ of \pageref{LastPage} \hfill%
      }
      \let\@evenfoot\@oddfoot
    }

    \AtBeginDocument{
        \pagestyle{headings}
        \thispagestyle{plain}
        \captionsetup{
            singlelinecheck=true  
        }
    }
    
    \renewcommand{\maketitle}{
      \begin{center}
        {\LARGE\bfseries
        A market-initialised accelerated failure time model for in-play football forecasting
        \par}
        \vspace{1em}
        {\large
        Lawrence Clegg$^{*}$\,\orcidlink{0009-0009-1292-4773},
        Zixing Song\,\orcidlink{0000-0002-8871-3990}, 
        John Cartlidge\,\orcidlink{0000-0002-3143-6355}
        \par}
        {\normalsize
        School of Engineering Mathematics and Technology, University of Bristol, UK
        \par}  
        {\normalsize
        \texttt{\{lawrence.clegg, zixing.song, john.cartlidge\}@bristol.ac.uk}
        \par}
        {\small
        $^{*}$Corresponding author
        \par}
        \vspace{1.5em}
        \begin{minipage}{0.9\textwidth}
        \small
        \noindent\textbf{Abstract---}\ignorespaces
        In-play football forecasting models are rarely compared with betting exchange prices, which aggregate information from many market participants. We make this comparison for several goal-arrival models, focusing on a Weibull accelerated failure time model with two extensions: initialising team strength parameters from Betfair Exchange prices at kick-off to capture pre-match market information, and including cumulative shots as a time-varying covariate to capture in-play information. The initialisation, which fits team-strength parameters to 1X2 and over/under prices by squared-error minimisation, applies to any intensity-based goal arrival model. Evaluated across 280 English Premier League matches at minute intervals, the market-initialised model remains measurably weaker than Betfair in-play prices (classification accuracy 68.5\% versus 69.5\%, ranked probability score 0.1325 versus 0.1281), but retains interpretable team-level parameters and covariate effects. A comparison with a market-implied Poisson model, a Cox process and a Bayesian birth process, all initialised from the same pre-match odds, confirms that market initialisation is the dominant driver of predictive accuracy. A betting simulation against Betfair in-play odds yields a 1.0\% return on investment over 37{,}467 bets (95\% interval $-3.8\%$ to $5.9\%$), which we regard as preliminary given the limited sample of matches and prices that may not have been executable.
        \vspace{1em}
        
        \noindent
        \textbf{Keywords:}
        betting markets; 
market calibration; 
goal arrival models; 
soccer; 
survival analysis; 
Weibull distribution

        \end{minipage}
        
        \vspace{2em}
      \end{center}
    }
    \makeatother
\fi

\begin{document}

\pdfpagewidth=8.5in
\pdfpageheight=11in
\setlength{\hoffset}{0in}
\setlength{\voffset}{0in}
\setlength{\oddsidemargin}{0in}
\setlength{\evensidemargin}{0in}
\setlength{\topmargin}{-0.5in}
\setlength{\headheight}{14pt}
\setlength{\headsep}{0.25in}
\setlength{\textwidth}{6.5in}
\setlength{\textheight}{9in}
\setlength{\hsize}{6.5in}
\setlength{\linewidth}{6.5in}
\setlength{\columnwidth}{6.5in}
\setlength{\marginparwidth}{0pt}
\setlength{\marginparsep}{0pt}

\ifnum\ARXIV=0
    \articletype{Research Article}
    \journalname{Journal of Quantitative Analysis in Sports}
    \journalyear{YYYY}
    \journalvolume{XX}
    \journalissue{X}
    \startpage{1}
    \aop
    \DOI{10.1515/sample-YYYY-XXXX}

    \title{A market-initialised accelerated failure time model for in-play football forecasting}
    \runningtitle{Market-initialised AFT model for in-play football forecasting}

\else
\fi

\ifnum\BLIND=1
    \author*[1]{Anonymous Author(s)}
    \runningauthor{Anonymous}
    \affil[1]{\protect\raggedright Institution withheld for blind review.}
\else
    \ifnum\ARXIV=0
        \author*[1]{Lawrence Clegg}
        \author[2]{Zixing Song}
        \author[3]{John Cartlidge}
        \runningauthor{Clegg et al.}
        \affil[1]{\protect\raggedright School of Engineering Mathematics and Technology, University of Bristol, Merchant Venturers Building, Woodland Road, Bristol, BS8 1UB, UK, E-mail: \href{mailto:lawrence.clegg@bristol.ac.uk}{lawrence.clegg@bristol.ac.uk}, \url{https://orcid.org/0009-0009-1292-4773}}
        \affil[2]{\protect\raggedright School of Engineering Mathematics and Technology, University of Bristol, Bristol, BS8 1UB, UK, \url{https://orcid.org/0000-0002-8871-3990}}
        \affil[3]{\protect\raggedright School of Engineering Mathematics and Technology, University of Bristol, Bristol, BS8 1UB, UK, \url{https://orcid.org/0000-0002-3143-6355}}
    \else
    \fi
\fi

\abstract{
In-play football forecasting models are rarely compared with betting exchange prices, which aggregate information from many market participants. We make this comparison for several goal-arrival models, focusing on a Weibull accelerated failure time model with two extensions: initialising team strength parameters from Betfair Exchange prices at kick-off to capture pre-match market information, and including cumulative shots as a time-varying covariate to capture in-play information. The initialisation, which fits team-strength parameters to 1X2 and over/under prices by squared-error minimisation, applies to any intensity-based goal arrival model. Evaluated across 280 English Premier League matches at minute intervals, the market-initialised model remains measurably weaker than Betfair in-play prices (classification accuracy 68.5\% versus 69.5\%, ranked probability score 0.1325 versus 0.1281), but retains interpretable team-level parameters and covariate effects. A comparison with a market-implied Poisson model, a Cox process and a Bayesian birth process, all initialised from the same pre-match odds, confirms that market initialisation is the dominant driver of predictive accuracy. A betting simulation against Betfair in-play odds yields a 1.0\% return on investment over 37{,}467 bets (95\% interval $-3.8\%$ to $5.9\%$), which we regard as preliminary given the limited sample of matches and prices that may not have been executable.
}
\maketitle

\section{Introduction}\label{sec:intro}\noindent
The prediction of association football match outcomes attracts considerable interest from the sports analytics and betting communities. In-play betting, where bettors use observed match events to update their forecasts during a match, now commands around 60\% of the sports betting market.\footnote{\url{https://www.mordorintelligence.com/industry-reports/online-sports-betting-market}} Betting exchanges such as Betfair Exchange (and prediction markets) have substantial in-play liquidity for elite-level football matches, with some bettors placing thousands of bets per match as they react to match events and market dynamics. Live data feeds from providers such as Stats Perform---whose Opta brand collects over 1 billion unique data points annually across more than 20 sports and 3{,}900 competitions\footnote{\url{https://www.statsperform.com/resource/opta-by-stats-perform-global-leader-ai-sports-data-analytics/}}---enable fully automated betting strategies in these markets. While association football is the most widely played and watched sport globally, forecasting match outcomes in-play presents clear challenges: the sport is inherently low-scoring, with around three goals per match on average, and matches can result in draws, creating a three-way outcome classification problem.

Pre-match forecasting has a long history, from early Poisson models \citep{maher1982modelling} to bivariate extensions \citep{dixon1997modelling, karlis2003analysis, boshnakov2017bivariate} and machine learning approaches \citep{bunker2024machine}. However, these methods do not address in-play updating. In higher-scoring sports, within-match updating is more straightforward: tennis models update win probabilities point-by-point \citep{klaassen2003forecasting, kovalchik2019calibration} and basketball models simulate possession-by-possession \citep{vstrumbelj2012simulating}. In football, goals are sufficiently rare that models must extract signal from non-scoring events to update forecasts meaningfully between goals.

A growing body of in-play work addresses this challenge. \citet{robberechts2021bayesian} model scoring intensity as a Poisson process conditioned on many match-event covariates, including expected threat, attacking passes, and duel success rates. Their pre-match baseline, however, rests on a single Elo rating differential, which captures relative strength but offers no team-specific attack or defence decomposition: two fixtures with the same Elo gap are assigned identical pre-match expected goals, regardless of the teams' scoring rates. Other notable point-process approaches \citep{dixon1998birth,volf2009random,maia2025stochastic,zou2020bayesian} model arrivals directly, but condition on score state, red cards, and time effects alone, drawing little signal from shot-level or event-derived data between goals.

\citet{leriou2025survival} model goal inter-arrival times directly, via a Weibull accelerated failure time model. The usage of an accelerated failure time model comes from survival analysis, where covariates accelerate or decelerate the waiting time to the event (goal). A gap-time formulation provides a natural basis for simulating match progressions. Their model achieves precise in-sample league reconstruction and its structure admits a straightforward extension to in-play covariates, which they leave as future work.

A further limitation shared by the above models is their inferior pre-match forecasting compared with betting markets, which have been consistently shown to be the best forecasters before kick-off \citep{hvattum2010using, wunderlich2025using, maia2025stochastic}. This likely constrains the in-play performance of these models from the outset, although in-play models have rarely been compared with in-play odds. For instance, both \citet{zou2020bayesian} and \citet{titman2015joint} bet against live market prices, but neither compares forecast accuracy to the market.  

Betting odds can be incorporated into goal models to close this gap. \citet{titman2015joint} replace team attack and defence parameters with covariates derived from pre-match bookmaker odds, which informs scoring rates but leaves no team-level parameters to interpret. Goal intensities have also been fitted to market prices by least squares, but to value bets rather than to forecast outcomes \citep{divos2020modelling, jottreau2009markets}. With forecasting as the aim, \citet{egidi2018combining} combine odds-implied and historical Poisson scoring rates with a per-match convex combination, achieving accuracy close to bookmaker odds and positive expected returns in the top four European leagues. However, the model forecasts only final scores, with no in-play updating.

Two limitations motivate the work we present. First, the Weibull gap-time framework of \citet{leriou2025survival} has not been applied in-play, despite the authors explicitly proposing this as future work and the structure admitting in-play covariates by design. Second, pre-match baselines estimated from historical data alone are substantially less accurate than market prices, yet no in-play forecasting model fits its team-strength parameters directly to them. 

We address both limitations by extending the Weibull accelerated failure time model with in-play covariates and pre-match market initialisation. We incorporate cumulative shots as a time-varying covariate, capturing attacking volume beyond what the score alone reflects. Rather than estimating team strength independently of the market, we initialise the model's scoring-rate parameters from Betfair Exchange prices at kick-off and use the model's structure for in-play updating. Although this step resembles model calibration in finance, where model parameters are fitted to observed prices, we call it market initialisation to distinguish it from probability calibration (the agreement between forecast probabilities and observed frequencies).

We evaluate all models against Betfair Exchange in-play prices across 280 matches from gameweek~11 onwards of the 2024/25 season, using log-loss, Ranked Probability Score, and classification accuracy at minute intervals. We compare the Weibull model against the Bayesian birth process of \citet{zou2020bayesian}, the Cox process of \citet{maia2025stochastic}, and three market-implied benchmarks. The market-initialised Weibull and Maia models are indistinguishable from a Poisson benchmark fitted to the same prices, while the Zou model lags, which indicates that initialisation is the dominant driver of forecasting performance. Market initialisation raises accuracy from 67.7\% to 68.5\%, against 69.5\% for Betfair, while the Weibull model retains interpretable team-level parameters and covariate effects. A betting simulation with the market-initialised Weibull model against Betfair in-play odds yields 1.0\% ROI with Kelly staking over 37{,}467 bets, which we regard as preliminary.

\subsection{Related work}\label{sec:related}\noindent
Considering pre-match forecasting, \citet{maher1982modelling} demonstrated that a Poisson regression model with team-specific attack and defence parameters could adequately describe match scores, establishing a framework for subsequent forecasting models. Notable extensions to this framework include incorporating time-varying team strengths with a low-score dependence correction \citep{dixon1997modelling}, using a bivariate Poisson model with diagonal inflation to better capture draws \citep{karlis2003analysis}, and implementing a Weibull count process with copula-induced dependence \citep{boshnakov2017bivariate}. 
Player data has been used to forecast match outcomes, with \citet{holmes2024forecasting} introducing a Skellam regression approach that uses player ratings with positional interactions. Machine learning methods have also been applied to predict outcomes, with gradient-boosted tree models demonstrating competitive performance \citep{bunker2024machine}.

Scoring rates are known to rise through the course of a match \citep{ayana2025temporal}, and trailing teams have an elevated scoring rate in the second half \citep{silva2016analysis}. \citet{dixon1998birth} first introduced a bivariate non-homogeneous Poisson birth process in which home and away goal rates depend on the current scoreline. \citet{volf2009random} extended this to a semi-parametric Cox model with a non-parametric baseline intensity for each team, estimated directly from observed goal times, allowing scoring rates to vary freely over match time rather than following a parametric trend. \citet{titman2015joint} took a different approach and jointly modelled goals and bookings as an eight-dimensional counting process under Weibull proportional hazards, finding that red cards sharply alter scoring rates but yellow cards do not, and that home scoring rates fall once the home team leads. Each of these models accounts for the evolving match state, yet assumes team-ability parameters remain fixed throughout.

Building on the birth process framework of \citet{dixon1998birth}, \citet{zou2020bayesian} update team scoring intensities in-play via conjugate Gamma posteriors as goals are observed, without consideration of the information contained in shots, cards, and other match events that might foreshadow goals.

\citet{robberechts2021bayesian} address this limitation by treating scoring intensity as a temporal stochastic process with regression weights that evolve throughout the match. Beyond score and time, they incorporate expected threat \citep{singh2019xT}, attacking passes, duel success rates, and cards, enabling win probability to respond to shifts in momentum before goals are scored. However, their use of Elo ratings as the pre-match baseline captures only relative team strength, precluding the matchup-specific attack--defence decomposition needed for informative scoreline forecasts.

Following the shift towards greater use of in-play covariates, \citet{klemp2021play} evaluated the incremental value of 18 performance indicators---including shots, passes, space control, and running distance---for match outcome prediction. Using an ordered logistic regression to forecast second-half results, they found that first-half goals added no significant predictive value for second-half outcomes once pre-match bookmaker odds were included.

More recently, \citet{maia2025stochastic} treat goal arrivals as Cox processes with dynamic regressors for score differential, red-card difference, a half-time indicator, and the log-ratio of pre-match market values. They find that a team's scoring intensity drops by over 30\% upon receiving a red card, and trailing teams exhibit 10--20\% higher intensity. Although they provide a general structure for incorporating in-play covariates into goal intensity models, they do not consider shot-based covariates.

\section{Data and evaluation}\label{sec:data}\noindent
We collect data from several sources for the four English Premier League seasons 2021/22--2024/25 inclusive. WhoScored.com records a variety of timestamped match events, from which we extract goal arrival times, red card events, and injury time added. For shot-level detail, we use FBref.com, which records every shot together with its expected goals (xG) and post-shot expected goals (PSxG) values.

Betfair Exchange freely provides minute-by-minute historical data for football markets, recording the last traded price for each outcome. Betfair suspends in-play markets after goals, penalties and red cards, clearing the order book \citep{croxson2014information}, so trades just before or after a suspension, or late in matches when liquidity is thin, can record unrepresentative prices. We treat prices that are at least 0.05 in implied probability away from both neighbouring minutes this way, and replace 110 such prices in 2024/25 with the preceding price.\footnote{Lower cutoffs than 0.05 worsen Betfair's RPS, suggesting they remove genuine price movements.}

Of the 1{,}520 scheduled fixtures, four with missing or incomplete data are excluded. The remaining 1{,}516 are split at gameweek~11 of the 2024/25 season, with the 1{,}236 earlier matches used for training and the 280 from gameweek~11 onwards for evaluation. We start at gameweek~11 so that every club, including the promoted sides, has at least ten league matches in the season. We cross-checked that all but three of the 39{,}590 shots are noted at the same minute in both the FBref and the WhoScored data. For the three disputed shots, we use the FBref time. Appendix~\ref{app:data} lists all consistency checks.

\subsection{Evaluation metrics}\label{sec:metrics}\noindent
We evaluate each model's performance using three complementary metrics over the predicted probability distribution $\mathbf{p} = (p_H, p_D, p_A)$. The Ranked Probability Score (RPS) measures the accuracy of cumulative probability forecasts \citep{epstein1969scoring}:
\begin{equation}
    \text{RPS} = \frac{1}{2} \sum_{i=1}^{2} \left( \sum_{j=1}^{i} p_j - \sum_{j=1}^{i} a_j \right)^2,
\end{equation}
where outcomes are ordered as home win, draw, away win, and $a_j = 1$ if outcome $j$ occurred, 0 otherwise. RPS is standard in football forecasting \citep[see][]{constantinou2012solving} because it penalises predictions further from the true outcome more heavily than those nearby. However, \citet{wheatcroft2021evaluating} argues this ``sensitivity to distance'' provides no practical benefit, finding that log-loss identifies superior forecasters more efficiently. Hence, we also report log-loss, $-\log p_y$ where $y \in \{H, D, A\}$ denotes the realised outcome. Unlike RPS, log-loss considers only the probability assigned to the outcome and penalises confident incorrect predictions severely. We also report classification accuracy, though this discards probabilistic information.

We evaluate forecasts at every minute throughout each match, from minute~0 (pre-match) to the final whistle as determined by the last recorded event. We report aggregate metrics across all evaluation points; we also plot log-loss by minute, revealing how forecast accuracy evolves as match information accumulates and match-time remaining decreases. Since minutes within a match are correlated, we report match-clustered confidence intervals from a bootstrap that resamples whole matches with replacement (10{,}000 resamples).

\section{A Weibull model for in-play prediction}\label{sec:model}\noindent
\subsection{Accelerated failure time model}\noindent
\citet{leriou2025survival} introduce a Weibull accelerated failure time (AFT) model \citep{kalbfleisch2011statistical} for goal inter-arrival times using match data from the English Premier League 2018/19 season. Rather than addressing ``How many goals will a team score?'', they use the time between successive goals to answer ``When will a goal be scored?'' such that in a match between home team $H$ and away team $A$, goal arrival times follow:
\begin{align}
T_H &\sim \text{Weibull}(\gamma, \lambda_H), \\
T_A &\sim \text{Weibull}(\gamma, \lambda_A).
\end{align}
For $\gamma>1$, the conditional scoring hazard increases with time since the last goal. \citet{leriou2025survival} estimate $\gamma=1.13$ with a 95\% posterior interval of [1.078, 1.186]. They also fit a model with a score-state-dependent shape parameter, 0.94 for the leading team, 1.04 when tied and 1.73 for the trailing team, which achieved a superior fit as measured by the Deviance Information Criterion (DIC $= 10{,}870$ vs.\ $11{,}010$).

We could not replicate this score-state pattern.\footnote{This is the only discrepancy we found. The intercept, home advantage and common shape of \citet{leriou2025survival} all replicate closely on our data, as reported at the end of this section.} Labelling each interval by the score that was held \emph{during} it, we obtain shape parameters of 1.18, 1.17 and 1.15 for the leading, tied and trailing team. Labelling each interval instead by the score \emph{after} the goal that ends it reproduces the pattern (0.94, 1.10 and 1.86) in the original study, but this label is set by the very goal whose timing is being modelled. To test whether the labelling alone creates the pattern, we simulated the 1{,}236 training fixtures from our fitted single-shape model, in which scoring does not depend on the score. Labels by the score after the goal again give 0.95, 1.07 and 1.84, while labels by the score during the interval give no effect (1.16, 1.21 and 1.14). The score-state effect of \citet{leriou2025survival} therefore appears to arise from labelling intervals by their outcome, and with intervals labelled by the score during them, a score-state shape fits worse than a single shape ($\Delta$BIC $= 16.1$). A separate shape for each half is also not preferred by BIC ($\Delta$BIC $= 2.0$).

\citet{heuer2012past} examine the same questions in over four decades of Bundesliga matches. They find that the conceding team's scoring rate dips by about 10\% after a goal and recovers within roughly ten minutes. Our shape of 1.17 implies a larger dip for both teams, with the hazard over the ten minutes after a goal averaging 29\% below its level at thirty minutes. They also find that the score has little effect on which team scores next, which supports a single shape over separate shapes for the leading, tied and trailing team.

To determine $\lambda_H$ and $\lambda_A$, the ``standard vanilla formulation'' of \citet{karlis2003analysis} is used:
\begin{align}
\log \mathbb{E}[T_H] &= \mu + \beta_{\text{home}} + a_H + d_A, \\
\log \mathbb{E}[T_A] &= \mu + a_A + d_H,
\end{align}
where $\mu$ is the intercept, $\beta_{\text{home}}$ captures home advantage, and each team $k$ has attacking ability $a_k$ and defensive ability $d_k$. Lower attacking parameters indicate a stronger attack (shorter expected time to score); higher defensive parameters indicate a stronger defence (longer expected time for opponent to score). For identifiability, sum-to-zero constraints are imposed: $\sum_k a_k = \sum_k d_k = 0$.

For the half-time interval, \citet{leriou2025survival} found that treating time as continuous rather than censoring and resetting at the 45th minute yielded a better fit (DIC $11{,}010$ vs.\ $11{,}070$). They also investigated dependence between goal arrival times using a Marshall--Olkin bivariate Weibull distribution and random effects models, but found no improvement over the independent Weibull model.

We estimate all parameters by maximum likelihood on the observed goal times, right-censored at the final whistle, in contrast to the Bayesian approach of \citet{leriou2025survival}, and use BIC for model comparison. We fit this model on 1{,}236 in-sample matches from four EPL seasons (2021/22--2024/25). The intercept $\mu = 4.21$ closely matches \citeauthor{leriou2025survival}'s estimate of 4.03, we obtain $\beta_{\text{home}} = -0.18$ compared to their $-0.19$, and the common shape is $\gamma = 1.17$ (SE 0.015) against their 1.13. We now describe our adaptations for in-play prediction.

\subsection{In-play simulation}\label{sec:inplay}\noindent
We generate in-play forecasts by Monte Carlo simulation: at each prediction point, we sample 10{,}000 match completions from the current minute to full time, conditioning on the observed score and elapsed time. Outcome probabilities are obtained from the proportions of home wins, draws, and away wins across simulations. We also apply several adaptations to the base model: time-decaying team strengths that evolve across gameweeks, a conditional sampling mechanism for the Weibull distribution, and in-play covariates that update the scoring rate as events occur. 

\subsection{Time decay}\noindent
The performance of a football team notoriously varies throughout a season due to tactical adjustments, managerial changes, injuries and transfers. While \citet{leriou2025survival}'s choice of static team strength parameters is appropriate for validation of the accelerated failure time approach within a single season, we use a larger dataset and hence re-estimate team parameters after each gameweek using all previous matches.

We implement exponential time decay to capture the evolution of team strengths. Specifically, at each gameweek, all previous matches are weighted by
\begin{equation}
w_k = \exp\left(-\xi \cdot \frac{d_k}{3.5}\right),
\end{equation}
where $d_k$ is the number of days elapsed since match $k$, and division by 3.5 converts to half-weeks.

We use $\xi = 0.0065$ per half-week following \citet{dixon1997modelling}, who optimised this value by maximising predictive log-likelihood on match outcomes. This yields a half-life of approximately one year. 

\subsection{Conditional Weibull}\noindent
To adapt the framework for in-play prediction, we first consider that the Weibull distribution is not memoryless: the probability of a goal in the next minute depends on how long the current scoreline has persisted. We must therefore condition on the playing time $s$ elapsed since the last goal (or kick-off), which runs on across half-time. Following \citet{kleinbaum2012survival}, we sample $T - s \mid T > s$ via inverse transform:
\begin{equation}\label{eq:conditional_weibull}
T_{\text{remaining}} = \left( s^\gamma + \frac{-\log(U)}{\lambda} \right)^{1/\gamma} - s, \quad U \sim \text{Uniform}(0,1).
\end{equation}

Since we simulate from the current minute to full time, we must estimate the match endpoint, which depends on stoppage time added by the referee at the end of each half. While \citet{leriou2025survival} sample total stoppage uniformly from $[3, 7]$ minutes, we use half-specific means computed exclusively from the 100 training matches preceding the evaluation period: 3.1 minutes for the first half and 6.4 for the second. We restrict this computation to the current season because the rules for added time changed twice. From 2023/24, referees added the exact time lost to goals, substitutions and other stoppages, following guidance from the International Football Association Board, and from 2024/25 the first 30 seconds after each goal are no longer added.\footnote{\url{https://www.premierleague.com/news/3617054} and \url{https://www.premierleague.com/news/4076251}} Second-half added time averaged 5.2 minutes in 2022/23 and 7.1 in 2023/24. Although stoppage time can be modelled from match incidents such as goals, substitutions and cards \citep{watanabe2015determinants}, the simple mean performs comparably to an oracle that knows the true match duration on our evaluation set (for the base Weibull model evaluated every five minutes, log-loss 0.749 vs.\ 0.747 and accuracy 66.2\% for both).

\subsection{Shot volume}\label{subsec:shotquality}

Goals are rare, so the scoreline is a coarse record of attacking output, whereas shots accumulate throughout a match. Let $t_{k,j}$ denote the minute of team $k$'s $j$-th shot, whether on or off target. Hence,
\begin{equation}\label{eq:cum_shots}
S_k(t) = \sum_{j:\, t_{k,j} \le t} 1
\end{equation}
is the running total  of shots from kick-off. Since $S_k(t)$ grows with $t$, we subtract an estimated linear baseline $\bar{S}(t)$, obtained by regressing $S_k(t)$ on $t$ across all team-matches in the 1{,}236 training matches. The shots deviation covariate is
\begin{equation}\label{eq:shots_dev}
\psi_k(t) = S_k(t) - \bar{S}(t).
\end{equation}
Positive values indicate above-typical shot volume for the match stage. At prediction minute $M$, $\psi_k(M)$ is held fixed across all Monte Carlo paths, so future shots are not simulated. Shots vary widely in scoring potential, and each can instead be weighted by its estimated scoring probability. This idea dates back to \citet{pollard1997measuring}, who modelled the probability of a goal by logistic regression on shot location and context, and the resulting measure is now termed expected goals (xG). Post-shot expected goals (PSxG) refines xG using where the shot ends up in the goal. It is defined only for shots on target, so off-target shots contribute zero. FBref records both measures for every shot (Section~\ref{sec:data}). We define analogous deviation covariates for cumulative goals, xG and PSxG, and compare all four in Section~\ref{subsec:covariates}.

\subsection{Covariates}\label{subsec:covariates}

\begin{table}[tb]
  \caption{Nested comparison of in-play covariates, 1{,}236 training matches.}
  \label{tab:cov_comparison}
  \ifnum\ARXIV=1 \centering \fi
  \begin{tabular}{lcccccc}
 & M0 & M1 & M2 & M3 & M4 & M5 \\
Added covariate & None & Red cards & Goals & PSxG & xG & Shots \\
\midrule
$\hat{\gamma}$ & $1.17$ & $1.17$ & $1.18$ & $1.17$ & $1.17$ & $1.17$ \\
 & $(0.012)$ & $(0.012)$ & $(0.013)$ & $(0.013)$ & $(0.012)$ & $(0.012)$ \\
$\hat{\beta}_{\text{red}}$ & --- & $-0.442$ & $-0.430$ & $-0.424$ & $-0.405$ & $-0.395$ \\
 &  & $(0.060)$ & $(0.060)$ & $(0.060)$ & $(0.060)$ & $(0.061)$ \\
$\hat{\beta}_{\text{cov}}$ & --- & --- & $-0.027$ & $-0.037$ & $-0.082$ & $-0.016$ \\
 &  &  & $(0.015)$ & $(0.020)$ & $(0.023)$ & $(0.004)$ \\
\midrule
$k$ & $1$ & $2$ & $3$ & $3$ & $3$ & $3$ \\
$\Delta$BIC & $0.0$ & $-40.8$ & $-35.5$ & $-35.8$ & $-45.0$ & $-51.5$ \\
LRT & --- & $49.3$ & $3.2$ & $3.4$ & $12.7$ & $19.2$ \\
  \end{tabular}

  \vspace{1mm}
  {\footnotesize Notes: M2--M5 each add one deviation covariate to M1. Standard errors in parentheses, conditional on team parameters fixed from the first estimation stage. $\beta_{\text{red}}$ is the coefficient of the red card difference and $\beta_{\text{cov}}$ that of the added covariate. $k$ counts estimated parameters and $\Delta$BIC is relative to M0 (BIC 37{,}352.6). LRT is the likelihood ratio test statistic of M1 against M0 and of M2--M5 against M1, each with one degree of freedom (5\% critical value 3.84).}
\end{table}

We add in-play covariates through the linear predictor and compare five candidates in Table~\ref{tab:cov_comparison}: the red card difference and the deviation covariates for goals, PSxG, xG and shots defined in Section~\ref{subsec:shotquality}. 

The log-expected time equations become
\begin{align}
\log \mathbb{E}[T_H] &= \mu + \beta_{\text{home}} + a_H + d_A + \boldsymbol{\beta}^\top \mathbf{x}_H
 \label{eq:cov_h} \\
\log \mathbb{E}[T_A] &= \mu + a_A + d_H + \boldsymbol{\beta}^\top \mathbf{x}_A \label{eq:cov_a}
\end{align}
where $\mathbf{x}_k$ is the vector of covariate values for team $k$ at the current prediction time, which in the final model holds the red card difference and the shots deviation $\psi_k$.

Red cards, though rare, substantially affect match dynamics. \citet{maia2025stochastic}, with 3{,}039 matches containing 715 red cards, report a 30\% reduction in goal intensity for a team at a player disadvantage. \citet{leriou2025survival} found only weak evidence given their smaller sample of 380 matches with 47 red cards. Our training set of 1{,}236 matches contains 157 red cards. We encode the red card difference as the opponent's red cards minus the team's own, so that positive values indicate a player advantage.

Table~\ref{tab:cov_comparison} reports a nested model comparison. Red cards yield a large improvement (M1), with the negative coefficient indicating that a team with a player advantage scores sooner on average. In the final model (M5), a one-player advantage shortens the expected time to the next goal by about a third. Among the deviation covariates, shots fit best (M5), ahead of xG (M4), while goals and PSxG do not improve on M1 once the extra parameter is penalised. PSxG averages 0.56 on scored shots against 0.20 on saved ones, so it partly restates the scoreline. Conditional on shots, xG adds nothing (LRT $= 0.5$), so we adopt the shots deviation. The shape parameters shift negligibly across specifications.

At each prediction minute $M$, we draw $N = 10{,}000$ Monte Carlo paths and take the win, draw, and loss proportions as the forecast. Algorithm~\ref{alg:simulation} describes the exact process, using remaining playing times calculated from the observed half and the stoppage-time estimates above.

\begin{algorithm}[tb]
\small
\caption{In-play Monte Carlo forecast at minute $M$ with $N = 10{,}000$ paths.}\label{alg:simulation}
\begin{algorithmic}[1]
\Require prediction minute $M$; number of paths $N$; score $(g_H, g_A)$;
  playing time $s$ since the last goal (or kick-off if none);
  log-expected goal-arrival times $\eta_k = \log \mathbb{E}[T_k]$, $k \in \{H,A\}$,
  from Equations~\eqref{eq:cov_h}--\eqref{eq:cov_a};
  shape $\gamma$;
  remaining playing time $R$ to the final whistle
\State $\lambda_k \gets
  \left(\Gamma(1 + 1/\gamma)\,\mathrm{e}^{-\eta_k}\right)^{\gamma}$,
  \; $k \in \{H,A\}$
\For{$i = 1, \ldots, N$}
    \State $(g_H^{(i)}, g_A^{(i)}, \tilde{s}, t_{\text{rem}}) \gets (g_H, g_A, s, R)$
    \While{$t_{\text{rem}} > 0$}
        \State Independently sample $\tau_k$ from $T_k - \tilde{s} \mid T_k > \tilde{s}$ via
          Equation~\eqref{eq:conditional_weibull}, for $k \in \{H, A\}$
        \State $\tau^* \gets \min(\tau_H, \tau_A)$
        \If{$\tau^* > t_{\text{rem}}$}\;\textbf{break}\EndIf
        \State $(g_H^{(i)}, g_A^{(i)}) \gets (g_H^{(i)}, g_A^{(i)}) +
          (\mathbf{1}[\tau_H \le \tau_A],\; \mathbf{1}[\tau_A < \tau_H])$
        \State $\tilde{s} \gets 0$; \quad $t_{\text{rem}} \gets t_{\text{rem}} - \tau^*$
          \Comment{only a goal resets the clock}
    \EndWhile
\EndFor
\State \Return
  $p_H = \tfrac{1}{N}\sum_i \mathbf{1}[g_H^{(i)} > g_A^{(i)}]$,\;
  $p_D = \tfrac{1}{N}\sum_i \mathbf{1}[g_H^{(i)} = g_A^{(i)}]$,\;
  $p_A = 1 - p_H - p_D$
\end{algorithmic}
\end{algorithm}

\subsection{Market-initialised team strengths}

\noindent
\citet{leriou2025survival} achieve an out-of-sample agreement rate of 70.5\% for binary outcomes (win-or-loss versus draw) when predicting the second half of the 2018/19 EPL season from the first, and reconstruct the same season's final league table in-sample with a root mean squared error of 1.4 positions. However, the betting market is widely recognised as the most accurate pre-match forecaster \citep{forrest2005odds, vstrumbelj2010online, wunderlich2025using}: on our evaluation set, the uninitialised Weibull achieves 52.9\% pre-match accuracy versus 56.8\% for Betfair (Table~\ref{tab:summary}). To analyse the model's effectiveness at forecasting in-play, we initialise the team strength parameters from market prices so that pre-match forecasts are approximately the same as those from Betfair.

To do this, we utilise the over/under markets where bettors wager on whether the total number of goals in a match exceeds a given threshold. For example, `over 2.5 goals' pays out if three or more goals are scored. These markets provide direct estimates of the goals distribution: prices at thresholds $g \in \{0.5, 1.5, 2.5, 3.5, 4.5\}$ give the market-implied probability $P(G > g)$ for total goals $G$, and collectively these five values help determine the probability of each scoreline when combined with the 1X2 market prices and the goal-arrival model. 

These additional probabilities resolve an identifiability issue: match outcome probabilities alone are consistent with many combinations of $(\lambda_H, \lambda_A)$ at different absolute scoring levels. Over/under thresholds fix the total scoring rate, yielding a unique solution. Denoting the log-expected scoring times $\eta_H = \mu + \beta_{\text{home}} + a_H + d_A$ and $\eta_A = \mu + a_A + d_H$, we find for each match the market-initialised parameters $(\eta_H^{\kappa}, \eta_A^{\kappa})$ that minimise
\begin{equation}\label{eq:calibration}
\mathcal{L} = \sum_{o \in \{H,D,A\}} \left(\hat{p}_o - p_o^{\text{mkt}}\right)^2 + \sum_{g \in \mathcal{G}} \left(\hat{p}_{>g} - p_{>g}^{\text{mkt}}\right)^2,
\end{equation}
where $\hat{p}_o$ and $\hat{p}_{>g}$ are the Weibull model's minute~0 probabilities at $(\eta_H, \eta_A)$ for match outcome $o$ and the over-$g$ goals market, and $\mathcal{G} = \{0.5, 1.5, 2.5, 3.5, 4.5\}$. The targets $p_o^{\text{mkt}}$ and $p_{>g}^{\text{mkt}}$ are the corresponding Betfair-implied probabilities, taken from the last prices traded before kick-off. Each evaluation of $\mathcal{L}$ estimates the model probabilities from 10{,}000 simulated paths, with common random numbers across evaluations. We minimise $\mathcal{L}$ with Powell's method \citep{powell1964efficient} initialised at the $(\eta_H, \eta_A)$ estimated by the model. Appendix~\ref{app:shifts} illustrates the resulting shifts for the first evaluation gameweek.

\section{Predictive performance}\label{sec:results}
\subsection{Comparison models}\label{sec:comparison}\noindent
We compare the Weibull model with two recent in-play models, three market-implied benchmarks and Betfair's in-play prices. The in-play models are the Bayesian birth process of \citet{zou2020bayesian} and the Cox process of \citet{maia2025stochastic}, which we refer to as the Zou and Maia models. We initialise both from Betfair Exchange prices at kick-off using the same objective as the Weibull model (Equation~\eqref{eq:calibration}), computing outcome probabilities by a recursive algorithm for the birth process and by Monte Carlo simulation for the Cox process. The market-implied benchmarks use the same prices with minimal structure, which separates the contribution of the prices from that of the model.

\subsubsection{Birth process with Bayesian updating}
\citet{zou2020bayesian} build on the non-homogeneous Poisson birth process of \citet{dixon1998birth}, where home and away goal intensities depend on team-specific attack ($\alpha$) and defence ($\beta$) parameters, score-state multipliers, and a home advantage factor. Their contribution is a Bayesian update of the composite scoring parameters $\theta_1 = \alpha_H \beta_A$ and $\theta_2 = \alpha_A \beta_H$ as goals are observed. Using Gamma priors conjugate to the Poisson likelihood, the posterior mean after observing $X(T)$ home goals by minute $T$ is
\begin{equation}
\hat{\theta}_1 = \frac{r_1 + X(T)}{r_1 + E_H(T)} \, \hat{\theta}_{01},
\end{equation}
where $\hat{\theta}_{01}$ is the prior estimate from historical matches, $E_H(T)$ is the expected home goals by minute $T$, and $r_1 = E_H(45)$ balances prior and in-match information: updating lightly in the first half and more aggressively in the second. Outcome probabilities are computed exactly via a recursive algorithm. Their update uses goal times alone.

Stoppage time is handled with injury-time multipliers that inflate the scoring rate at minutes 45 and 90, rather than simulating additional match time. To control for the addition of shots in our Weibull model, we multiply both intensities by $\exp(\xi_{\psi}\,\psi_k(t))$, with $\xi_{\psi}$ estimated jointly with the other parameters. The shots observed up to minute $T$ then enter $E_H(T)$ as a known multiplier, like the score-state and stoppage multipliers, so the update keeps its closed form. Both `Zou' variants reported below include this covariate.

\subsubsection{Cox process with dynamic regressors}
\citet{maia2025stochastic} model goal arrivals as Cox processes with time-varying regressors. Home goal intensity at minute $t$ is
\begin{equation}
\lambda_H(t) = \alpha_H \beta_A \exp\!\left(\delta + \xi_{\text{half}} \, h(t) + \xi_{\text{gd}} \, \Delta_H(t) + \xi_{\text{rc}} \, R_H(t)\right),
\end{equation}
where $\delta$ captures home advantage, $h(t) = 1$ if $t$ is in the second half and 0 otherwise, $\Delta_H(t)$ and $R_H(t)$ are the goal and red-card differences from team $H$'s perspective. The model jointly estimates goal, red card, and stoppage time processes. Red card arrivals follow a non-homogeneous Poisson process with power-law intensity, and stoppage times for each half are Poisson-distributed with log-linear regressors for red cards and goals scored, with the second half additionally including a close-match indicator. Outcome probabilities are computed via Monte Carlo simulation. 

In place of their market-value regressor, for which we lack comparable data, we initialise from pre-match odds. This modification was proposed by \citet{maia2025stochastic}. Their model with the regressor trails bookmaker odds before kick-off (RPS 0.2090 against 0.2006, their Table~9), whereas the initialised version is level with Betfair at minute~0 (Table~\ref{tab:summary}). To control for the addition of shots in our Weibull model, we extend the Cox process intensity with the per-team shots deviation covariate $\psi_k(t)$ defined in Equation~\eqref{eq:shots_dev}, applied to home and away intensities with a shared coefficient $\xi_{\psi}$. This coefficient is estimated jointly with the other parameters and held constant during forward simulation from the current prediction minute. Both `Maia' variants reported below include this covariate.

\subsubsection{Betfair Exchange}
The market benchmark uses Betfair Exchange 1X2 last-traded prices, normalised to implied probabilities. Since event times and exchange prices are recorded at one-minute resolution on separate clocks, prices at minute $M$ may not yet reflect a goal at minute $M$. Additionally, Betfair suspends in-play markets after goals (Section~\ref{sec:data}), which can lead to an unfair representation in the last traded price. We therefore use prices from minute $M + 2$, by which time the market has reacted to 99\% of goals in the 100 training matches of 2024/25 (97\% by $M + 1$). The shift favours the market: without it, Betfair's aggregate RPS would be 0.1320 rather than 0.1281. The prices used are also recent, 96\% having been traded within a minute of the evaluation minute. Forecasts at minute~0 are the one exception: we instead take the last price traded before kick-off, so that the market benchmark and the market-initialised models hold the same information.

\subsubsection{Market-implied benchmarks}
We use three benchmarks that take their team strengths from the same pre-match prices and add minimal in-play structure. First, $\poik$ is an independent Poisson model fitted by Equation~\eqref{eq:calibration}, whose expected remaining goals scale with the fraction of match time left \citep{dixon1998birth}. Second, $\poikt$ replaces this fraction with the empirical share of goals still to come in the training matches, which captures the rising scoring rate and stoppage-time excess reported by \citet{dixon1998birth}. Third, $\naivek$ is a multinomial logistic regression on pre-match log-odds, goal difference and minutes remaining, with no goal process.

\subsection{Results}\label{sec:accuracy}\noindent
We denote the base survival model as Weibull, and use $\wbp$ when extended with the shots covariate $\psi$. A subscript $\kappa$ indicates team strengths initialised from the last Betfair Exchange prices before kick-off; thus $\wbpc$ denotes the full model. The same convention applies to the comparison models: $\zoup$, $\zouc$, $\maiap$ and $\maiac$. The market-implied benchmarks are $\poik$, $\poikt$ (where the subscript $t$ marks a time-varying scoring rate) and $\naivek$. Betfair denotes odds-implied probabilities from the 1X2 market.

\begin{table}[tb]
\caption{Predictive performance before kick-off and across all evaluation minutes.}
\label{tab:summary}
\ifnum\ARXIV=1 \centering \fi
\begin{tabular}{lrrrrrr}
 & \multicolumn{3}{c}{Pre-match} & \multicolumn{3}{c}{All minutes} \\
\cmidrule(lr){2-4}\cmidrule(lr){5-7}
Model & Accuracy & RPS & Log-loss & Accuracy & RPS & Log-loss \\
\midrule
Weibull   & 0.529 & 0.2102 & 0.9985 & 0.677 & 0.1365 & 0.7213 \\
$\wbp$    & 0.529 & 0.2102 & 0.9987 & 0.677 & 0.1358 & 0.7184 \\
$\wbpc$   & 0.554 & \textbf{0.2008} & 0.9747 & \textbf{0.685} & \textbf{0.1325} & \textbf{0.7063} \\
\addlinespace
$\zoup$   & 0.507 & 0.2168 & 1.0185 & 0.670 & 0.1444 & 0.7675 \\
$\zouc$   & \textbf{0.568} & 0.2011 & \textbf{0.9740} & 0.680 & 0.1395 & 0.7492 \\
\addlinespace
$\maiap$  & 0.504 & 0.2168 & 1.0187 & 0.678 & 0.1377 & 0.7239 \\
$\maiac$  & 0.557 & 0.2017 & 0.9753 & \textbf{0.685} & 0.1329 & 0.7068 \\
\addlinespace
$\poik$   & \textbf{0.568} & 0.2011 & 0.9744 & \textbf{0.685} & 0.1332 & 0.7084 \\
$\poikt$  & \textbf{0.568} & 0.2011 & 0.9744 & \textbf{0.685} & 0.1332 & 0.7084 \\
$\naivek$ & 0.561 & 0.2044 & 0.9901 & 0.677 & 0.1381 & 0.7401 \\
\addlinespace
Betfair   & 0.568 & 0.2012 & 0.9755 & 0.695 & 0.1281 & 0.6831 \\
\end{tabular}

\smallskip
{\footnotesize Notes: 280 matches with 27{,}710 evaluated minutes. Subscript $\kappa$ indicates market-initialised variants and $\psi$ the shots covariate of Equation~\eqref{eq:shots_dev}. Bold marks the best value in each column excluding Betfair, including ties. Note that, by construction, $\poik$ and $\poikt$ have identical pre-match predictive scores, since the time-dependent adjustment is not applied before kick-off.
}
\end{table}

\begin{figure}[tb]
\centering
\includegraphics[width=0.75\textwidth]{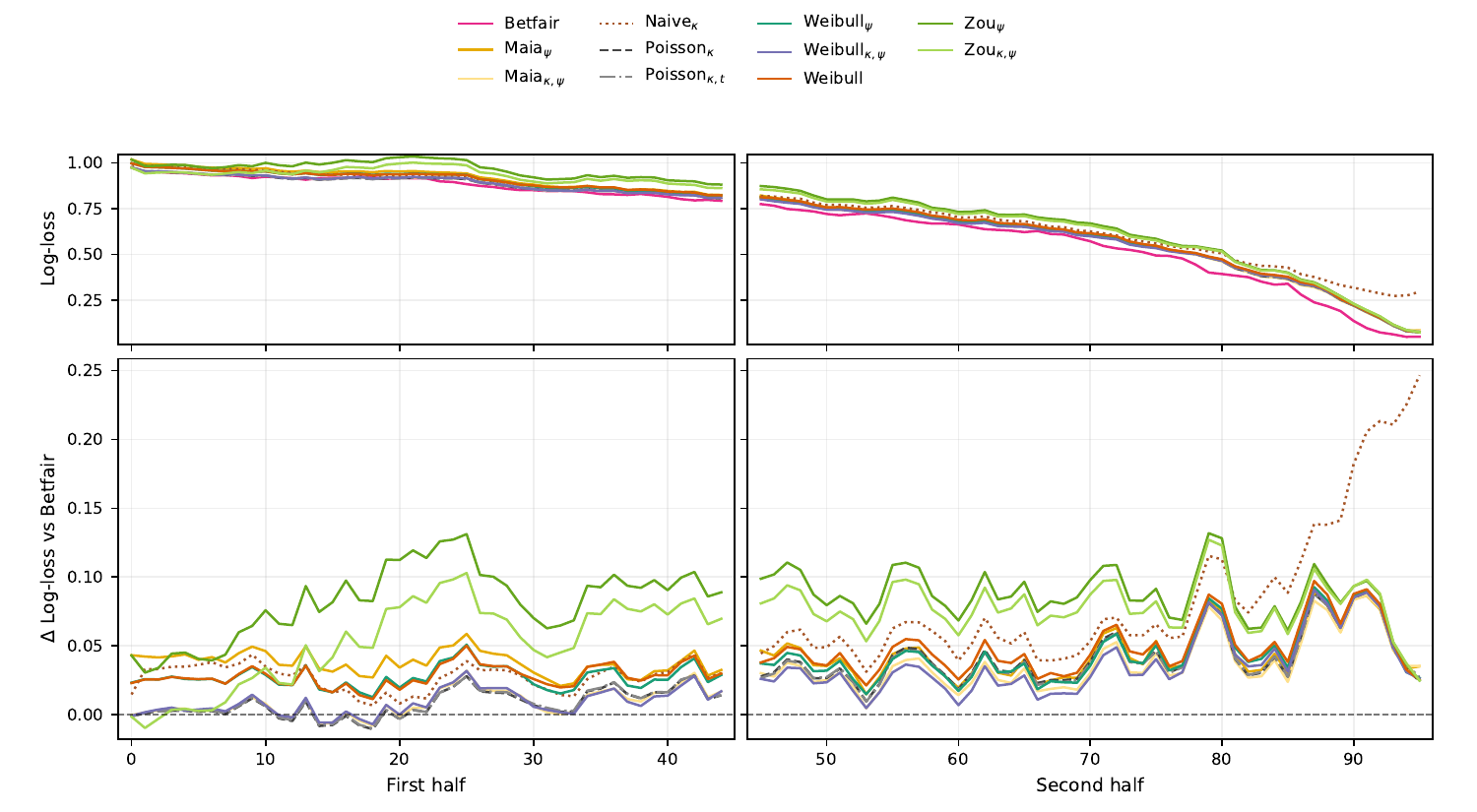}
\caption{Log-loss by match minute. Left: first half (minutes 0--44). Right: second half (minutes 45--95). Top: log-loss at each evaluation minute. Bottom: log-loss minus that of Betfair.}
\label{fig:logloss}
\end{figure}

At minute~0, all three market-initialised models ($\wbpc$, $\zouc$, $\maiac$) closely resemble Betfair (Table~\ref{tab:summary}, pre-match columns), confirming that market initialisation successfully incorporates market information into team-strength parameters across structurally different model classes, while retaining each model's structure for in-play updating. The Poisson benchmarks sit in the same range, while $\naivek$ is weaker at 0.990. We do not interpret these small RPS and log-loss advantages over Betfair as a real edge: sampling variation easily covers the differences at $N = 280$, and Betfair would likely outperform market-initialised models at scale.

Across all minutes, $\wbpc$ achieves a classification accuracy of 68.5\% against 69.5\% for Betfair, and Betfair retains the best RPS and log-loss. Betfair's advantage over $\wbpc$ is 1.0 percentage points in accuracy [0.3, 1.8], 0.0044 in RPS [0.0026, 0.0062] and 0.023 in log-loss [0.013, 0.034], with match-clustered 95\% intervals in brackets, so the market remains better on every metric. When both models incorporate shots and market initialisation, $\maiac$ slightly trails $\wbpc$ on RPS (0.1329 vs. 0.1325) and log-loss (0.7068 vs. 0.7063), while $\zouc$ lags substantially despite market initialisation. $\naivek$ is clearly worse (log-loss 0.7401), but the Poisson benchmarks $\poik$ and $\poikt$ are close to both (0.7084), so most of the accuracy comes from the pre-match prices rather than the goal-arrival structure.
The two Poisson benchmarks also achieve identical all-minutes predictive scores to the reported precision, suggesting that accounting for empirical time variation in scoring rates provides little additional predictive benefit.

Figure~\ref{fig:logloss} traces performance by match minute. Both $\wbpc$ and $\maiac$ perform close to Betfair. Betfair holds general dominance throughout the match as the market incorporates information beyond goals and shots that our covariates do not capture. Betfair's relative performance improves against all models at minute 45 (the second-half restart), as the half-time interval allows the market to fully absorb first-half information. $\wbp$ remains consistently above the market-initialised version, confirming the importance of market-initialised team strength estimates. The log-loss of $\zouc$ rises sharply around minute 20, as its Bayesian updating overrides the market-initialised pre-match strengths based on sparse goal evidence. In a 0--0 match, the update reduces scoring rates by 30\% at minute 20 and 50\% at half-time, pushing predictions heavily toward draws.

Table~\ref{tab:summary} separates the sources of these differences. In the Weibull model the shots covariate lowers aggregate RPS by 0.0007, while market initialisation lowers it by 0.003--0.005 in every architecture. Once initialised, the Weibull and Cox models are indistinguishable (difference 0.0004), but both improve on $\zouc$ by about 0.007.

\begin{figure}[tb]
\centering
\includegraphics[width=0.4\textwidth]{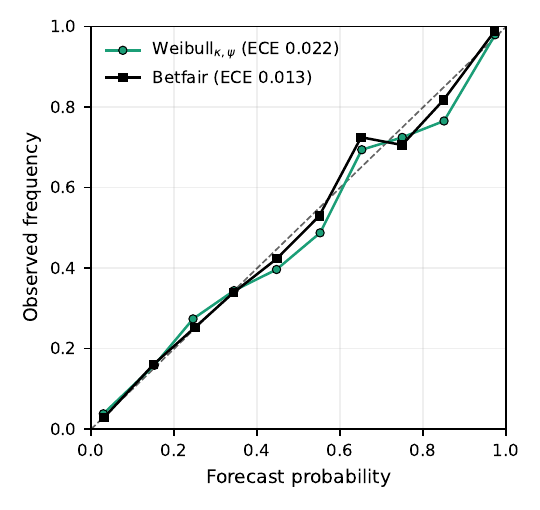}
\caption{Reliability diagram for $\wbpc$ and Betfair over the 1X2 probabilities at all 27{,}710 prediction points, ten equal-width bins. Legend values are the expected calibration error (ECE). The dashed line marks perfect calibration.}
\label{fig:reliability}
\end{figure}

Figure~\ref{fig:reliability} compares calibration. Both forecasts lie close to the diagonal below a probability of 0.4, but for $\wbpc$ forecasts between 0.4 and 0.6 and between 0.8 and 0.9, outcomes occur less often than predicted, so its expected calibration error (ECE), the bin-weighted mean absolute gap between forecast probability and observed frequency over all three outcomes, is higher than the market's (0.022 against 0.013). Appendix~\ref{app:calibration} shows the same diagram for every model.

\section{Betting simulation}\label{sec:betting}\noindent

\begin{table}[tb]
\caption{Betting simulation against Betfair in-play odds.}
\label{tab:betting}
\ifnum\ARXIV=1 \centering \fi
\begin{tabular}{llrrrrr}
Model & Strategy & Bets & Win (\%) & Net Profit & ROI (\%) & Sharpe \\
\midrule
$\wbp$ & Fixed payout & 36{,}969 & 39 & $144.59$ & $1.0$ & $0.38$ \\
        & Kelly & 36{,}969 & 39 & $-93.14$ & $-1.9$ & $-0.57$ \\
\addlinespace
$\wbpc$ & Fixed payout & 37{,}467 & 44 & $227.97$ & $1.4$ & $0.64$ \\
        & Kelly & 37{,}467 & 44 & $45.26$ & $1.0$ & $0.46$ \\
\addlinespace
$\maiac$ & Fixed payout & 35{,}408 & 44 & $-45.51$ & $-0.3$ & $-0.13$ \\
        & Kelly & 35{,}408 & 44 & $-23.04$ & $-0.5$ & $-0.25$ \\
\addlinespace
$\zouc$ & Fixed payout & 34{,}922 & 45 & $483.47$ & $3.2$ & $1.51$ \\
        & Kelly & 34{,}922 & 45 & $234.86$ & $2.5$ & $0.92$ \\
\addlinespace
$\poik$ & Fixed payout & 36{,}805 & 44 & $119.59$ & $0.7$ & $0.35$ \\
        & Kelly & 36{,}804 & 44 & $-0.67$ & $0.0$ & $-0.01$ \\
\end{tabular}

\smallskip
{\footnotesize Notes: 280 matches, with bets placed at every evaluation minute. Net profit is in units after 2\% commission on net match winnings, and ROI is net profit over total stake. Sharpe is the mean net profit per match over its standard deviation, annualised by $\sqrt{380}$ (matches per season).}
\end{table}

\noindent
As a secondary check, each model trades against Betfair in-play 1X2 odds at every evaluation minute (Table~\ref{tab:betting}). We compare two staking strategies, the Kelly criterion \citep{kelly1956new} and the fixed-payout staking of \citet{wilkens2021sports}. For model probability $p$ and Betfair decimal odds $o$, both back every outcome with positive expected value ($p o - 1 > 0$) and differ only in the stake. Fixed-payout staking stakes $1/o$, so every winning bet returns one unit, while Kelly staking stakes the fraction $f = (p o - 1)/(o - 1)$ of a one-unit bankroll. Following \citet{boshnakov2017bivariate}, the ``bankroll'' is reset before each bet, so stakes do not compound. A fractional Kelly stake therefore scales every bet equally and leaves ROI and the Sharpe ratio unchanged. We also deduct a 2\% commission on net winnings in each match, matching Betfair's ``BASIC'' plan.

\begin{figure}[tb]
\centering
\includegraphics[width=0.8\textwidth]{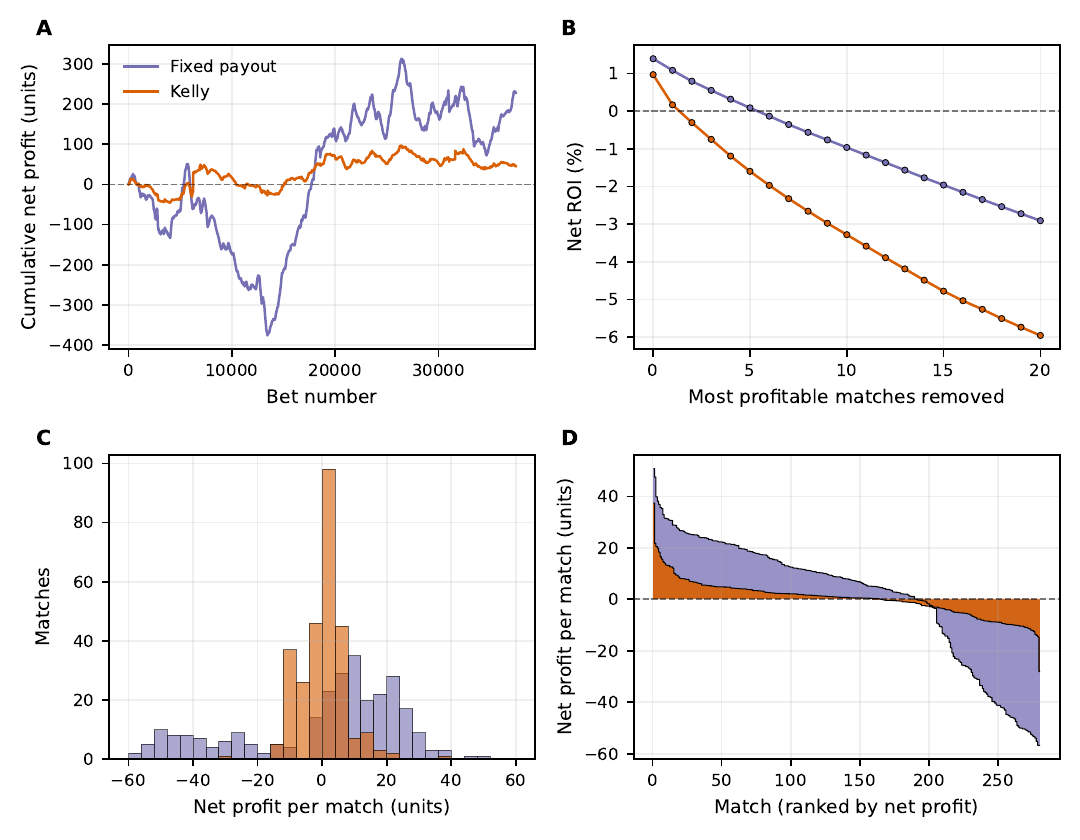}
\caption{Betting returns of $\wbpc$ against Betfair in-play odds, 280 matches, under fixed-payout and Kelly staking. (A) Cumulative net profit by bet. (B) Net ROI after removing the $k$ most profitable matches. (C) Distribution of net profit per match. (D) Net profit per match, matches ranked from most to least profitable. Commission of 2\% applied to net match winnings.}
\label{fig:bet_fragile}
\end{figure}

$\wbpc$ returns $1.4\%$ under fixed-payout staking and $1.0\%$ under Kelly staking (Figure~\ref{fig:bet_fragile}A). Bets within a match are strongly dependent, and the match-clustered 95\% intervals, $[-3.6\%, 6.2\%]$ and $[-3.8\%, 5.9\%]$, both include zero. The returns are also concentrated. The model profits in 68\% of matches under fixed-payout staking and 59\% under Kelly staking (Figure~\ref{fig:bet_fragile}C and D), but removing the six most profitable of the 280 matches (fixed payout) or the two most profitable (Kelly) makes the return negative (Figure~\ref{fig:bet_fragile}B). Neither return reflects better calibration, as $\wbpc$ has a higher expected calibration error than the market (0.022 against 0.013, Figure~\ref{fig:reliability}), and the market-implied $\poik$ earns $0.7\%$ under fixed-payout staking. Moreover, $\zouc$, among the least accurate models (Table~\ref{tab:summary}), earns the highest returns ($3.2\%$ fixed payout, $2.5\%$ Kelly). Last-traded prices may also not be executable, particularly around goals when the market is suspended. We therefore regard these results as preliminary: returns and forecast accuracy measure different things \citep{wunderlich2020betting}, and a profitability claim would need executable order-book prices and a larger sample.

\begin{figure}[tb]
\centering
\includegraphics[width=0.8\textwidth]{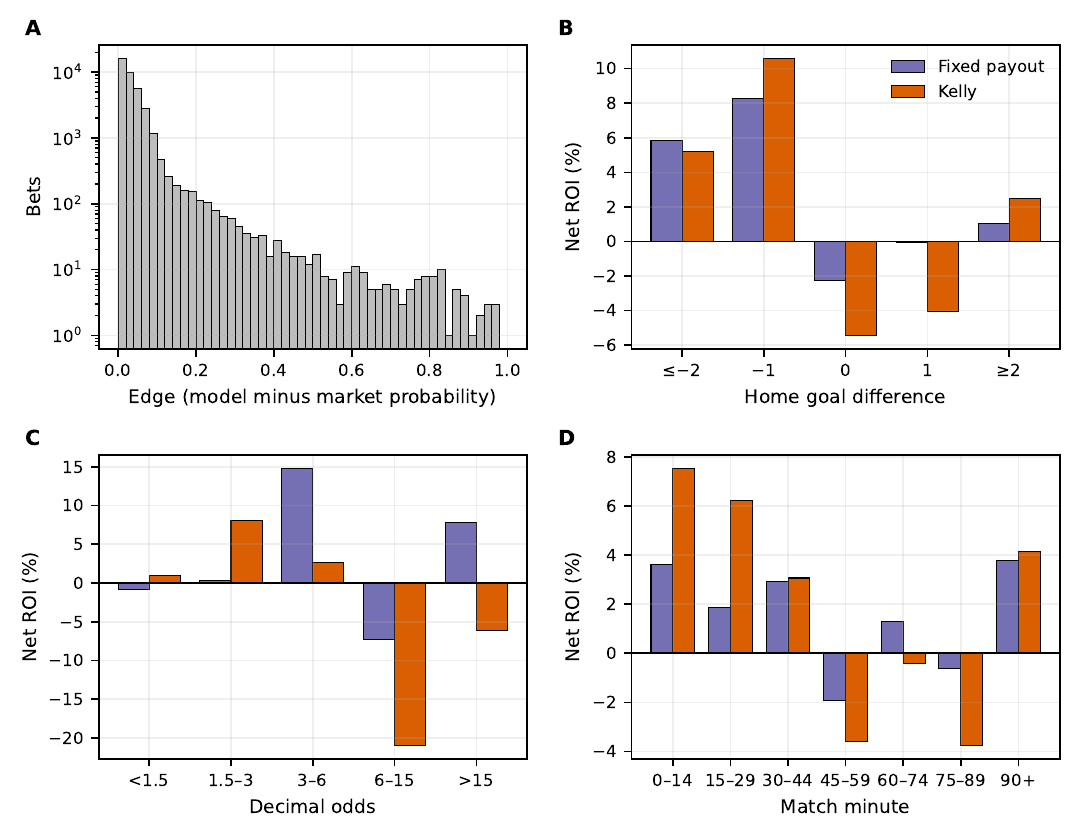}
\caption{Bets placed by $\wbpc$ against Betfair in-play odds, 280 matches. (A) Edge (model minus market probability) of the 37{,}467 bets placed, log scale. (B)--(D) Net ROI by home goal difference, decimal odds and match minute at the time of the bet, under fixed-payout (purple) and Kelly (orange) staking.}
\label{fig:bet_breakdown}
\end{figure}

Figure~\ref{fig:bet_breakdown} shows where the returns arise. Most bets carry an edge below 0.05, with a thin tail up to 1 (Figure~\ref{fig:bet_breakdown}A). Half of the bets (19{,}075) are placed at level scores, where both strategies lose, and the returns come from states in which the home side trails (Figure~\ref{fig:bet_breakdown}B). In those states the stake is mostly on the away side, and 2024/25 had the second-highest number of away wins in Premier League history \citep{bateman2025arsenal}. By odds, both strategies lose on outcomes priced between 6 and 15 (Figure~\ref{fig:bet_breakdown}C). By minute, both profit throughout the first half and lose between minutes 45 and 59 (Figure~\ref{fig:bet_breakdown}D).

\section{Conclusion}\label{sec:conclusion}\noindent
We have extended the Weibull accelerated failure time model of \citet{leriou2025survival} to in-play football forecasting, and examined the effect of initialising team strengths from Betfair Exchange prices at kick-off. The model remains less accurate than Betfair in-play prices on every metric, and a market-implied Poisson model initialised from the same prices performs similarly, so the accuracy of the model is mostly attributable to the imported market information. However, its interpretable team-level parameters and covariate effects remain valuable, as does an initialisation method that applies to any intensity-based goal arrival model. Our findings support the view of \citet{forrest2005odds} and \citet{wunderlich2025using} that betting odds encode information difficult to replicate from historical match data, and suggest that integrating market prices into goal arrival models is a productive direction for in-play forecasting. The betting simulation yielded small positive returns under both staking strategies, but given the limited sample of matches and the assumed fulfilment of orders, no conclusion can be drawn on in-play market efficiency.

These results carry limitations. First, the betting returns rest on Betfair's last-traded prices, which may differ from the prices executable at the time of prediction. Second, the 280 matches come from a single season. \citet{winkelmann2024betting} show by simulation that samples of this length can produce apparent inefficiency in fully efficient markets by chance alone, and \citet{wunderlich2020betting} show that positive returns can arise without a superior forecasting model. We therefore regard the accuracy results as more robust than the betting returns. Finally, FBref has since lost access to the Opta statistics underlying our shot records.\footnote{\url{https://www.sports-reference.com/blog/2026/01/fbref-stathead-data-update/}}

Future research could extend the model in several directions. The market's consistent advantage from minute 30 onwards, and particularly at the second-half restart, suggests the model would benefit from stronger in-play covariates. Substitutions are a natural candidate: managers use them to alter team shape and attacking intent, and they are easily observed. Expected threat from passing sequences \citep{singh2019xT} could capture sustained momentum more effectively than shot-level statistics, which are sparse.

Extending the evaluation to multiple leagues and seasons would test the generalisability of both the underlying Weibull model and the initialisation approach, particularly in leagues where second-half scoring patterns or market liquidity may differ. Access to Betfair's Pro tier historical data (or a similar feed from a prediction market), which records limit order book depth at 50\,ms intervals, would allow executable rather than last-traded prices to be used, supporting a more rigorous test of betting profitability.

A final direction is to optimise the model directly for profitability rather than for likelihood. \citet{wunderlich2026does} show theoretically that the equivalence of accuracy- and profitability-optimal model selection rests on restrictive assumptions, and \citet{hubavcek2023beating} demonstrate in NBA betting that an accuracy-inferior model can produce systematic profits when decorrelating predictions from market prices is included in the optimisation function. While our initialisation step targets pre-match market consistency, team-strength estimation could instead target profit, fitting parameters that maximise expected returns against historical in-play prices rather than market fit at kick-off.

\begin{funding}
\ifnum\BLIND=1
    Funding details withheld for blind review.
\else
    LC's PhD is supported by a studentship from the Engineering and Physical Sciences Research Council (EPSRC) Doctoral Training Partnership (DTP), grant number EP/W524414/1. JC is supported by the UK Research and Innovation (UKRI) Engineering and Physical Sciences Research Council (EPSRC), grant number EP/Y028392/1: AI for Collective Intelligence (AI4CI). ZS received research funding from the School of Engineering Mathematics and Technology, University of Bristol, which supported his research activities during the period in which this work was undertaken.
\fi
\end{funding}

\ifnum\ARXIV=1
    \bibliographystyle{abbrvnat} 
\else
    \bibliographystyle{chicago}
\fi
\ifnum\BBL=1
    
\else
    \bibliography{references}

\begin{thebibliography}{}

\bibitem[\protect\citeauthoryear{Ayana, Ehlert, Ehlert, Santagata, Torricelli,
  and Klein}{Ayana et~al.}{2025}]{ayana2025temporal}
Ayana, G., A.~Ehlert, J.~Ehlert, L.~Santagata, M.~Torricelli, and B.~Klein
  (2025).
\newblock Temporal dynamics of goal scoring in soccer.
\newblock arXiv preprint arXiv:2501.18606.
  \url{https://arxiv.org/abs/2501.18606}.

\bibitem[\protect\citeauthoryear{Bateman}{Bateman}{2025}]{bateman2025arsenal}
Bateman, R. (2025).
\newblock Arsenal need to rediscover home comforts to succeed in 2025-26.
\newblock Opta Analyst,
  \url{https://theanalyst.com/articles/arsenal-home-form-concern-ahead-of-2025-26}.
\newblock Accessed: 2026-10-08.

\bibitem[\protect\citeauthoryear{Boshnakov, Kharrat, and McHale}{Boshnakov
  et~al.}{2017}]{boshnakov2017bivariate}
Boshnakov, G., T.~Kharrat, and I.~G. McHale (2017).
\newblock A bivariate {W}eibull count model for forecasting association
  football scores.
\newblock {\em International Journal of Forecasting\/}~{\em 33}, 458--466.

\bibitem[\protect\citeauthoryear{Bunker, Yeung, and Fujii}{Bunker
  et~al.}{2024}]{bunker2024machine}
Bunker, R., C.~Yeung, and K.~Fujii (2024).
\newblock Machine learning for soccer match result prediction.
\newblock In {\em Artificial Intelligence, Optimization, and Data Sciences in
  Sports}, pp.\  7--49. Springer.

\bibitem[\protect\citeauthoryear{Constantinou and Fenton}{Constantinou and
  Fenton}{2012}]{constantinou2012solving}
Constantinou, A.~C. and N.~E. Fenton (2012).
\newblock Solving the problem of inadequate scoring rules for assessing
  probabilistic football forecast models.
\newblock {\em Journal of Quantitative Analysis in Sports\/}~{\em 8}, 1--14.

\bibitem[\protect\citeauthoryear{Croxson and Reade}{Croxson and
  Reade}{2014}]{croxson2014information}
Croxson, K. and J.~J. Reade (2014).
\newblock Information and efficiency: {G}oal arrival in soccer betting.
\newblock {\em The Economic Journal\/}~{\em 124\/}(575), 62--91.

\bibitem[\protect\citeauthoryear{Divos}{Divos}{2020}]{divos2020modelling}
Divos, P. (2020).
\newblock {\em Modelling of the In-Play Football Betting Market}.
\newblock Ph.\ D. thesis, University College London.

\bibitem[\protect\citeauthoryear{Dixon and Robinson}{Dixon and
  Robinson}{1998}]{dixon1998birth}
Dixon, M. and M.~Robinson (1998).
\newblock A birth process model for association football matches.
\newblock {\em Journal of the Royal Statistical Society: Series D (The
  Statistician)\/}~{\em 47}, 523--538.

\bibitem[\protect\citeauthoryear{Dixon and Coles}{Dixon and
  Coles}{1997}]{dixon1997modelling}
Dixon, M.~J. and S.~G. Coles (1997).
\newblock Modelling association football scores and inefficiencies in the
  football betting market.
\newblock {\em Journal of the Royal Statistical Society: Series C (Applied
  Statistics)\/}~{\em 46}, 265--280.

\bibitem[\protect\citeauthoryear{Egidi, Pauli, and Torelli}{Egidi
  et~al.}{2018}]{egidi2018combining}
Egidi, L., F.~Pauli, and N.~Torelli (2018).
\newblock Combining historical data and bookmakers' odds in modelling football
  scores.
\newblock {\em Statistical Modelling\/}~{\em 18}, 436--459.

\bibitem[\protect\citeauthoryear{Epstein}{Epstein}{1969}]{epstein1969scoring}
Epstein, E.~S. (1969).
\newblock A scoring system for probability forecasts of ranked categories.
\newblock {\em Journal of Applied Meteorology (1962-1982)\/}~{\em 8}, 985--987.

\bibitem[\protect\citeauthoryear{Forrest, Goddard, and Simmons}{Forrest
  et~al.}{2005}]{forrest2005odds}
Forrest, D., J.~Goddard, and R.~Simmons (2005).
\newblock Odds-setters as forecasters: The case of {E}nglish football.
\newblock {\em International Journal of Forecasting\/}~{\em 21}, 551--564.

\bibitem[\protect\citeauthoryear{Heuer and Rubner}{Heuer and
  Rubner}{2012}]{heuer2012past}
Heuer, A. and O.~Rubner (2012).
\newblock How does the past of a soccer match influence its future? {C}oncepts
  and statistical analysis.
\newblock {\em PLoS ONE\/}~{\em 7\/}(11), e47678.

\bibitem[\protect\citeauthoryear{Holmes and McHale}{Holmes and
  McHale}{2024}]{holmes2024forecasting}
Holmes, B. and I.~G. McHale (2024).
\newblock Forecasting football match results using a player rating based model.
\newblock {\em International Journal of Forecasting\/}~{\em 40}, 302--312.

\bibitem[\protect\citeauthoryear{Hub{\'a}{\v{c}}ek and
  {\v{S}}{\'\i}r}{Hub{\'a}{\v{c}}ek and
  {\v{S}}{\'\i}r}{2023}]{hubavcek2023beating}
Hub{\'a}{\v{c}}ek, O. and G.~{\v{S}}{\'\i}r (2023).
\newblock Beating the market with a bad predictive model.
\newblock {\em International Journal of Forecasting\/}~{\em 39}, 691--719.

\bibitem[\protect\citeauthoryear{Hvattum and Arntzen}{Hvattum and
  Arntzen}{2010}]{hvattum2010using}
Hvattum, L.~M. and H.~Arntzen (2010).
\newblock Using {ELO} ratings for match result prediction in association
  football.
\newblock {\em International Journal of Forecasting\/}~{\em 26\/}(3), 460--470.

\bibitem[\protect\citeauthoryear{Jottreau}{Jottreau}{2009}]{jottreau2009markets}
Jottreau, B. (2009).
\newblock How do markets play soccer?
\newblock Working paper, Universit\'e Paris-Est.

\bibitem[\protect\citeauthoryear{Kalbfleisch and Prentice}{Kalbfleisch and
  Prentice}{2011}]{kalbfleisch2011statistical}
Kalbfleisch, J.~D. and R.~L. Prentice (2011).
\newblock {\em The Statistical Analysis of Failure Time Data\/} (2nd ed.).
\newblock Hoboken, NJ: John Wiley \& Sons.

\bibitem[\protect\citeauthoryear{Karlis and Ntzoufras}{Karlis and
  Ntzoufras}{2003}]{karlis2003analysis}
Karlis, D. and I.~Ntzoufras (2003).
\newblock Analysis of sports data by using bivariate {P}oisson models.
\newblock {\em Journal of the Royal Statistical Society: Series D (The
  Statistician)\/}~{\em 52}, 381--393.

\bibitem[\protect\citeauthoryear{Kelly}{Kelly}{1956}]{kelly1956new}
Kelly, J.~L. (1956).
\newblock A new interpretation of information rate.
\newblock {\em The Bell System Technical Journal\/}~{\em 35}, 917--926.

\bibitem[\protect\citeauthoryear{Klaassen and Magnus}{Klaassen and
  Magnus}{2003}]{klaassen2003forecasting}
Klaassen, F.~J. and J.~R. Magnus (2003).
\newblock Forecasting the winner of a tennis match.
\newblock {\em European Journal of Operational Research\/}~{\em 148}, 257--267.

\bibitem[\protect\citeauthoryear{Kleinbaum and Klein}{Kleinbaum and
  Klein}{2012}]{kleinbaum2012survival}
Kleinbaum, D.~G. and M.~Klein (2012).
\newblock {\em Survival Analysis: A Self-Learning Text\/} (3rd ed.).
\newblock Springer.

\bibitem[\protect\citeauthoryear{Klemp, Wunderlich, and Memmert}{Klemp
  et~al.}{2021}]{klemp2021play}
Klemp, M., F.~Wunderlich, and D.~Memmert (2021).
\newblock In-play forecasting in football using event and positional data.
\newblock {\em Scientific Reports\/}~{\em 11}, 24139.

\bibitem[\protect\citeauthoryear{Kovalchik and Reid}{Kovalchik and
  Reid}{2019}]{kovalchik2019calibration}
Kovalchik, S. and M.~Reid (2019).
\newblock A calibration method with dynamic updates for within-match
  forecasting of wins in tennis.
\newblock {\em International Journal of Forecasting\/}~{\em 35}, 756--766.

\bibitem[\protect\citeauthoryear{Leriou and Ntzoufras}{Leriou and
  Ntzoufras}{2025}]{leriou2025survival}
Leriou, I. and I.~Ntzoufras (2025).
\newblock Survival modeling of goal arrival times in {E}nglish {P}remier
  {L}eague.
\newblock {\em Computational Statistics\/}~{\em 40}, 2109--2133.

\bibitem[\protect\citeauthoryear{Maher}{Maher}{1982}]{maher1982modelling}
Maher, M.~J. (1982).
\newblock Modelling association football scores.
\newblock {\em Statistica Neerlandica\/}~{\em 36}, 109--118.

\bibitem[\protect\citeauthoryear{Maia, Pennanen, da~Silva, and Targino}{Maia
  et~al.}{2026}]{maia2025stochastic}
Maia, L. F.~G., T.~Pennanen, M.~A. da~Silva, and R.~S. Targino (2026).
\newblock Stochastic modelling of football matches using dynamic regressors.
\newblock {\em International Journal of Forecasting\/}~{\em 42}, 181--202.

\bibitem[\protect\citeauthoryear{Murphy}{Murphy}{1973}]{murphy1973new}
Murphy, A.~H. (1973).
\newblock A new vector partition of the probability score.
\newblock {\em Journal of Applied Meteorology\/}~{\em 12\/}(4), 595--600.

\bibitem[\protect\citeauthoryear{Pollard and Reep}{Pollard and
  Reep}{1997}]{pollard1997measuring}
Pollard, R. and C.~Reep (1997).
\newblock Measuring the effectiveness of playing strategies at soccer.
\newblock {\em Journal of the Royal Statistical Society Series D: The
  Statistician\/}~{\em 46}, 541--550.

\bibitem[\protect\citeauthoryear{Powell}{Powell}{1964}]{powell1964efficient}
Powell, M.~J. (1964).
\newblock An efficient method for finding the minimum of a function of several
  variables without calculating derivatives.
\newblock {\em The Computer Journal\/}~{\em 7}, 155--162.

\bibitem[\protect\citeauthoryear{Robberechts, Van~Haaren, and
  Davis}{Robberechts et~al.}{2021}]{robberechts2021bayesian}
Robberechts, P., J.~Van~Haaren, and J.~Davis (2021).
\newblock A {B}ayesian approach to in-game win probability in soccer.
\newblock In {\em Proceedings of the 27th ACM SIGKDD Conference on Knowledge
  Discovery \& Data Mining}, pp.\  3512--3521.

\bibitem[\protect\citeauthoryear{Silva and Swartz}{Silva and
  Swartz}{2016}]{silva2016analysis}
Silva, R.~M. and T.~B. Swartz (2016).
\newblock Analysis of substitution times in soccer.
\newblock {\em Journal of Quantitative Analysis in Sports\/}~{\em 12},
  113--122.

\bibitem[\protect\citeauthoryear{Singh}{Singh}{2019}]{singh2019xT}
Singh, K. (2019).
\newblock Introducing {E}xpected {T}hreat ({xT}).
\newblock Accessed: 2026-05-15.

\bibitem[\protect\citeauthoryear{{\v{S}}trumbelj and
  Robnik-{\v{S}}ikonja}{{\v{S}}trumbelj and
  Robnik-{\v{S}}ikonja}{2010}]{vstrumbelj2010online}
{\v{S}}trumbelj, E. and M.~Robnik-{\v{S}}ikonja (2010).
\newblock Online bookmakers' odds as forecasts: The case of {E}uropean soccer
  leagues.
\newblock {\em International Journal of Forecasting\/}~{\em 26}, 482--488.

\bibitem[\protect\citeauthoryear{{\v{S}}trumbelj and
  Vra{\v{c}}ar}{{\v{S}}trumbelj and
  Vra{\v{c}}ar}{2012}]{vstrumbelj2012simulating}
{\v{S}}trumbelj, E. and P.~Vra{\v{c}}ar (2012).
\newblock Simulating a basketball match with a homogeneous {M}arkov model and
  forecasting the outcome.
\newblock {\em International Journal of Forecasting\/}~{\em 28}, 532--542.

\bibitem[\protect\citeauthoryear{Titman, Costain, Ridall, and Gregory}{Titman
  et~al.}{2015}]{titman2015joint}
Titman, A., D.~Costain, P.~Ridall, and K.~Gregory (2015).
\newblock Joint modelling of goals and bookings in association football.
\newblock {\em Journal of the Royal Statistical Society Series A: Statistics in
  Society\/}~{\em 178}, 659--683.

\bibitem[\protect\citeauthoryear{Volf}{Volf}{2009}]{volf2009random}
Volf, P. (2009).
\newblock A random point process model for the score in sport matches.
\newblock {\em IMA Journal of Management Mathematics\/}~{\em 20}, 121--131.

\bibitem[\protect\citeauthoryear{Watanabe, Wicker, and Reuter}{Watanabe
  et~al.}{2015}]{watanabe2015determinants}
Watanabe, N.~M., P.~Wicker, and J.~C. Reuter (2015).
\newblock Determinants of stoppage time awarded to teams in the {E}nglish
  {P}remier {L}eague.
\newblock {\em International Journal of Sport Finance\/}~{\em 10}, 310--327.

\bibitem[\protect\citeauthoryear{Wheatcroft}{Wheatcroft}{2021}]{wheatcroft2021evaluating}
Wheatcroft, E. (2021).
\newblock Evaluating probabilistic forecasts of football matches: the case
  against the ranked probability score.
\newblock {\em Journal of Quantitative Analysis in Sports\/}~{\em 17},
  273--287.

\bibitem[\protect\citeauthoryear{Wilkens}{Wilkens}{2021}]{wilkens2021sports}
Wilkens, S. (2021).
\newblock Sports prediction and betting models in the machine learning age: The
  case of tennis.
\newblock {\em Journal of Sports Analytics\/}~{\em 7\/}(2), 99--117.

\bibitem[\protect\citeauthoryear{Winkelmann, {\"O}tting, Deutscher, and
  Makarewicz}{Winkelmann et~al.}{2024}]{winkelmann2024betting}
Winkelmann, D., M.~{\"O}tting, C.~Deutscher, and T.~Makarewicz (2024).
\newblock Are betting markets inefficient? {E}vidence from simulations and real
  data.
\newblock {\em Journal of Sports Economics\/}~{\em 25\/}(1), 54--97.

\bibitem[\protect\citeauthoryear{Wunderlich}{Wunderlich}{2025}]{wunderlich2025using}
Wunderlich, F. (2025).
\newblock Using the wisdom of crowds in sports: how performance analysis in
  football can benefit from the information enclosed in betting odds.
\newblock {\em International Journal of Performance Analysis in Sport\/}~{\em
  25}, 687--706.

\bibitem[\protect\citeauthoryear{Wunderlich, Caparr{\'o}s, and
  Memmert}{Wunderlich et~al.}{2026}]{wunderlich2026does}
Wunderlich, F., M.~G. Caparr{\'o}s, and D.~Memmert (2026).
\newblock Does the consideration of market prices in model selection increase
  model profitability? {E}vidence from theory, artificial data and real-world
  data.
\newblock {\em International Journal of Forecasting\/}.
\newblock In press, corrected proof.

\bibitem[\protect\citeauthoryear{Wunderlich and Memmert}{Wunderlich and
  Memmert}{2020}]{wunderlich2020betting}
Wunderlich, F. and D.~Memmert (2020).
\newblock Are betting returns a useful measure of accuracy in (sports)
  forecasting?
\newblock {\em International Journal of Forecasting\/}~{\em 36}, 713--722.

\bibitem[\protect\citeauthoryear{Zou, Song, and Shi}{Zou
  et~al.}{2020}]{zou2020bayesian}
Zou, Q., K.~Song, and J.~Shi (2020).
\newblock A {B}ayesian in-play prediction model for association football
  outcomes.
\newblock {\em Applied Sciences\/}~{\em 10}, 2904.

\end{thebibliography}
\fi

\appendix
\section{Data consistency checks}\label{app:data}\noindent
Table~\ref{tab:sanity} lists the data checks applied before model estimation. Three fixtures lack a WhoScored event feed altogether, and the feed for Newcastle United v Tottenham (17 October 2021) stops in the first half. These four fixtures are excluded. FBref and WhoScored agree on every team's shot count in the remaining matches. WhoScored counts minutes from 0 and FBref from 1, and after allowing for this offset only three shots are timed differently, each by one minute. In the Betfair data, rows that do not quote all three outcomes are dropped, and one-minute price spikes are replaced with the preceding price (Section~\ref{sec:data}). Apart from these price spikes, no check finds an error in the evaluation matches.

\begin{table}[tb]
\caption{Data consistency checks across the 2021/22--2024/25 seasons.}
\label{tab:sanity}
\ifnum\ARXIV=1 \centering \fi
\small
\begin{tabular}{llrl}
\toprule
Check & Checked & Flagged & Action \\
\midrule
WhoScored event feed exists & 1{,}520 fixtures & 3 & Excluded \\
Event feed reaches the final whistle & 1{,}517 fixtures & 1 & Excluded \\
Final score agrees with FBref & 1{,}517 fixtures & 1 & Same fixture as above \\
Shot count agrees for both teams & 1{,}516 fixtures & 0 & -- \\
Shot minute agrees & 39{,}590 shots & 3 & FBref minute used \\
Shot has an xG value & 39{,}590 shots & 0 & -- \\
Added time at most 15 minutes per half & 1{,}516 fixtures & 6 & Retained, timestamps agree \\
Goal event not duplicated & 4{,}501 goals & 0 & -- \\
Betfair quotes all three outcomes (2024/25) & 42{,}877 in-play minutes & 3 & Dropped \\
No one-minute price spike (2024/25) & 42{,}874 in-play prices & 110 & Preceding price used \\
Kick-off 1X2 and over/under prices exist & 280 evaluation & 0 & -- \\
\bottomrule
\end{tabular}

\smallskip
{\footnotesize Notes: Checked is the number of items examined and Flagged the number failing the check. Betfair checks cover 2024/25, the season whose in-play prices enter the evaluation.}
\end{table}
\section{Calibration of all models}\label{app:calibration}
\begin{table}[tb]
\caption{Calibration of all models.}
\label{tab:calibration}
\ifnum\ARXIV=1 \centering \fi
\begin{tabular}{lrrrrr}
\toprule
Model & Brier & Reliability & Resolution & ECE & Slope \\
\midrule
Weibull & 0.1401 & 0.00046 & 0.0820 & 0.014 & 0.90 \\
$\wbp$ & 0.1397 & 0.00062 & 0.0826 & 0.018 & 0.89 \\
$\wbpc$ & 0.1374 & 0.00095 & 0.0848 & 0.022 & 0.90 \\
$\maiap$ & 0.1407 & 0.00041 & 0.0814 & 0.014 & 0.90 \\
$\maiac$ & 0.1375 & 0.00086 & 0.0847 & 0.021 & 0.90 \\
$\zoup$ & 0.1464 & 0.00332 & 0.0776 & 0.045 & 0.68 \\
$\zouc$ & 0.1429 & 0.00302 & 0.0812 & 0.044 & 0.69 \\
$\poik$ & 0.1375 & 0.00092 & 0.0847 & 0.020 & 0.89 \\
$\poikt$ & 0.1375 & 0.00088 & 0.0846 & 0.020 & 0.90 \\
$\naivek$ & 0.1433 & 0.00070 & 0.0788 & 0.020 & 0.91 \\
\addlinespace
Betfair & 0.1330 & 0.00047 & 0.0886 & 0.013 & 1.01 \\
\bottomrule
\end{tabular}

\smallskip
{\footnotesize Notes: 1X2 probabilities at all 27{,}710 prediction points, ten equal-width bins. Brier is the mean squared error of the pooled outcome probabilities and approximately equals reliability $-$ resolution $+$ uncertainty (0.2222 for every model). Lower reliability and higher resolution are better. Slope is the coefficient on the forecast log-odds in a logistic recalibration, and a slope below 1 indicates overconfidence.}
\end{table}

\begin{figure}[tb]
\centering
\includegraphics[width=0.9\textwidth]{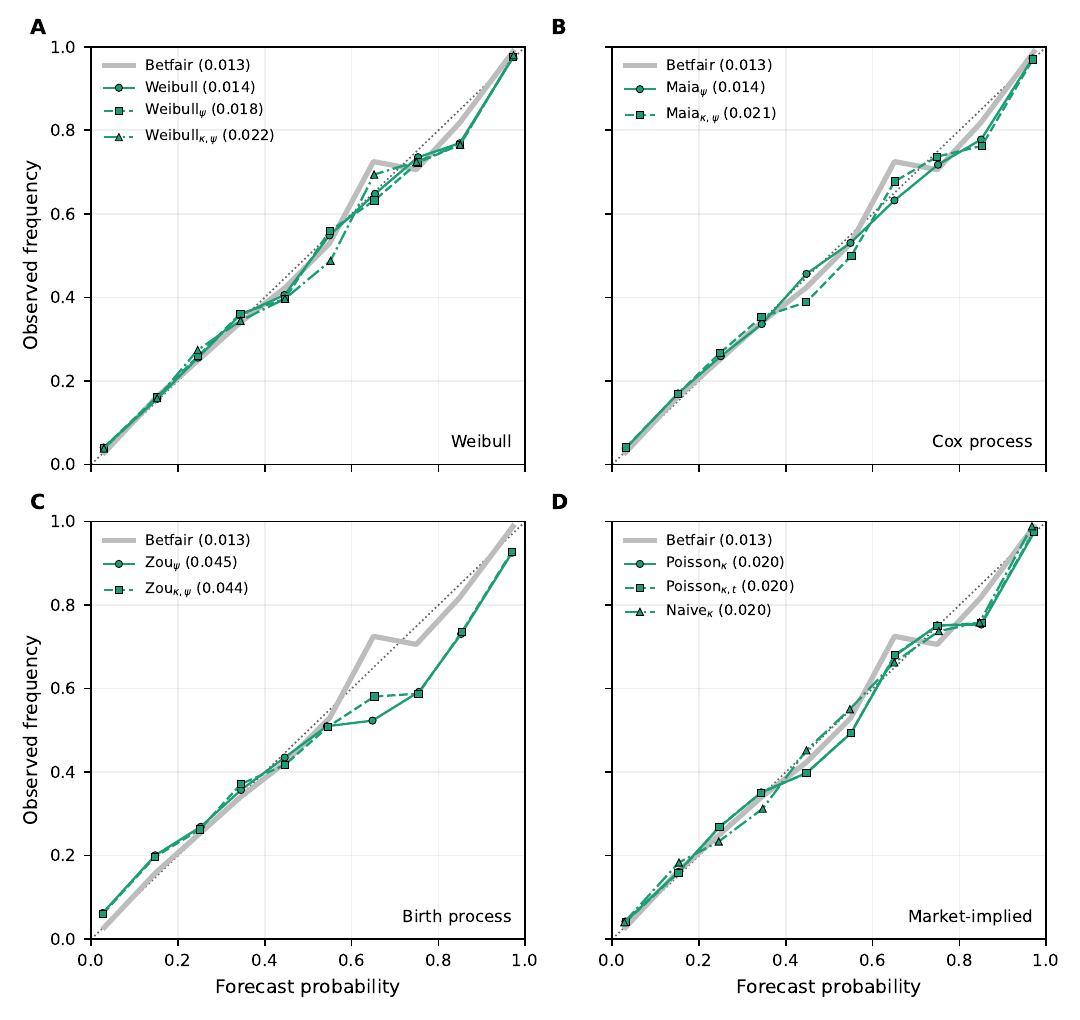}
\caption{Reliability diagrams for all models over the 1X2 probabilities at all 27{,}710 prediction points, ten equal-width bins. (A) Weibull, (B) Cox process, (C) birth process, (D) market-implied benchmarks. The thick grey line is Betfair in every panel and the dotted line marks perfect calibration. Legend values are the expected calibration error (ECE).}
\label{fig:reliability_all}
\end{figure}

\noindent Figure~\ref{fig:reliability_all} extends Figure~\ref{fig:reliability} to every model in Table~\ref{tab:summary}. The uninitialised Weibull and Cox models are close to Betfair (ECE 0.014 for Weibull, 0.018 for $\wbp$ and 0.014 for $\maiap$, against 0.013). Market initialisation raises the ECE to 0.022 for $\wbpc$ and 0.021 for $\maiac$, close to the market-implied benchmarks (0.020). Table~\ref{tab:calibration} shows why initialisation still improves accuracy. In the decomposition of the Brier score of \citet{murphy1973new}, initialisation leads to a large gain in the resolution of $\wbp$ from 0.0826 to 0.0848, a larger gain than its loss in reliability (0.00062 to 0.00095). Apart from the birth process, the model-based forecasts are slightly overconfident (calibration slope about 0.9, against 1.01 for Betfair). The birth-process models are the worst calibrated, with slopes of 0.68 and 0.69 and ECE of 0.045 and 0.044: they overstate outcomes forecast between 0.6 and 0.8 and understate those forecast below 0.2.

\section{Market initialisation shifts}\label{app:shifts}
\begin{figure}[tb]
\centering
\includegraphics[width=0.6\textwidth]{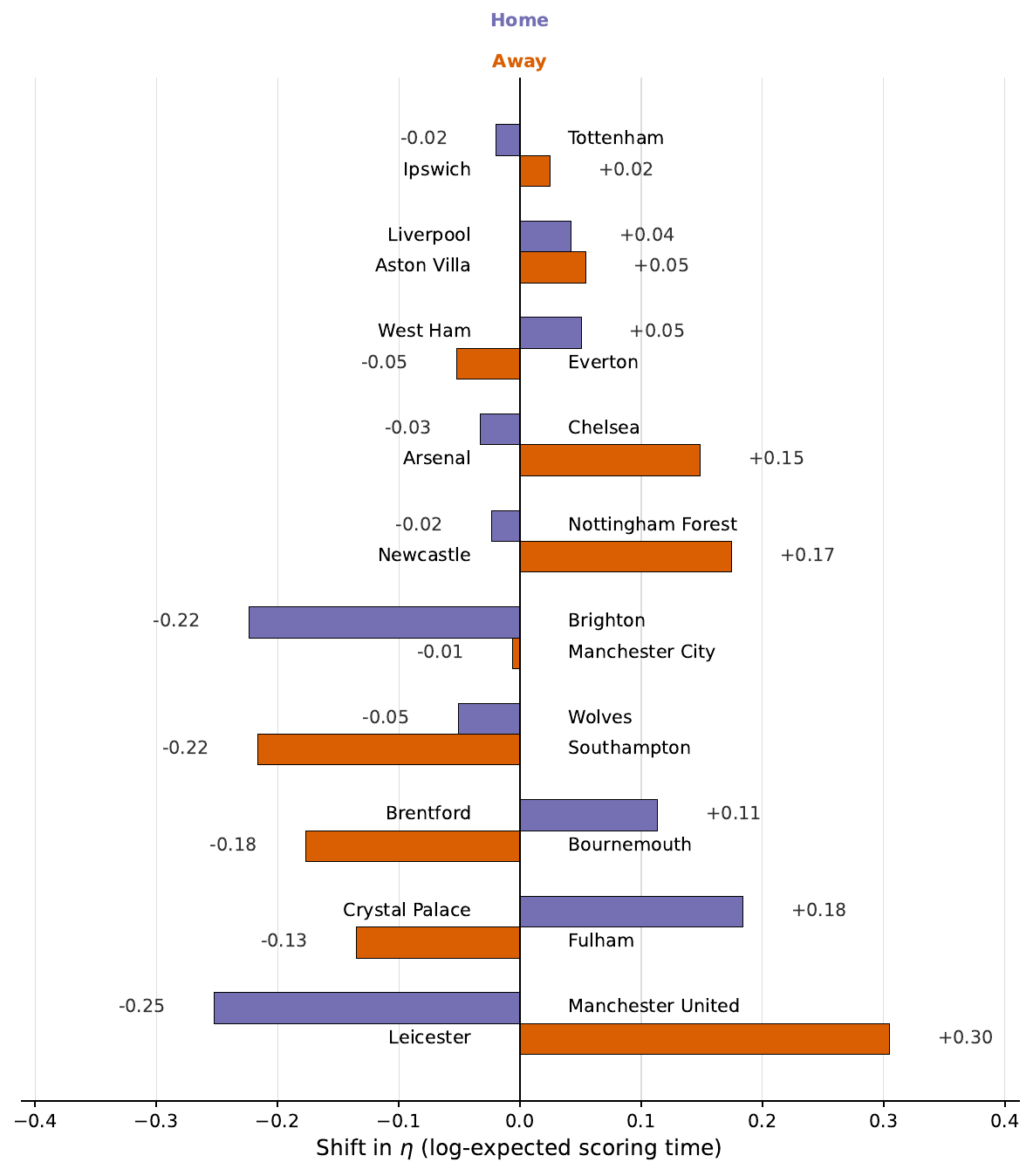}
\caption{Market initialisation shifts to team scoring parameters for gameweek~11 (first evaluation gameweek, 10 matches). Each paired row shows the shift from uninitialised to market-initialised $\eta$ for the home (purple) and away (orange) team. Positive values indicate the market assigns a slower scoring rate than the model. Rows ordered by total absolute shift.}
\label{fig:calibration_shift}
\end{figure}

\noindent
Figure~\ref{fig:calibration_shift} illustrates the initialisation shifts for gameweek~11 of the 2024/25 season. Shifts range from near zero to 0.30 (a 26-minute increase in Leicester's expected time to score against Manchester United), reflecting the information gap between the model's scoreline-derived parameters and the market's richer assessment of team quality. The shifts are positive and negative in roughly equal number (9 of 20 positive), so the market reallocates scoring strength between teams more than it changes the overall scoring rate.

\end{document}